\documentstyle[preprint,tighten,floats,prd,aps]{revtex}
\include {epsf}
     
\def\ra{\rightarrow}          
\def\bar{\overline}          
          
\def\a{\alpha}  
\def\b{\beta}          
          
\def\d{\delta}          
          
\def\m{\mu}          
          
\def\e{\epsilon}          
          
\def\th{\theta}

\def\bar{\overline}

\def\eV{{\rm eV}}   
\def\be{\begin{equation}}  
\def\ee{\end{equation}}  
\def\bea{\begin{eqnarray}}  
\def\eea{\end{eqnarray}}         
\def\D{\Delta}
\begin{document}          
\setcounter{page}{0}          
\thispagestyle{empty}          
\topskip 2.5  cm          
\topskip 0.  cm          
\begin{flushright}          
Lund-Mph-98/10     
\end{flushright}          
\vspace{0.3 cm}          
\centerline{\Large\bf Zee Mass Matrix and Bi-Maximal Neutrino Mixing}  
\vskip 1 cm          
\begin{center} 
{\bf Cecilia Jarlskog 
\footnote{E-mail address: cecilia.jarlskog@matfys.lth.se}${}^a$, 
Masahisa Matsuda \footnote{Permanent address:Department 
of Physics and Astronomy, 
Aichi University of Education, Kariya, Aichi 448, Japan. 
E-mail address:mmatsuda@matfys.lth.se}${}^a$ 
\vskip 0.5 cm } 
{\bf Solveig Skadhauge 
\footnote{E-mail address: solveig@matfys.lth.se}${}^a$ 
and 
Morimitsu Tanimoto 
\footnote{E-mail address: tanimoto@edserv.ed.ehime-u.ac.jp}${}^b$} 
\end{center} 
\centerline{${}^a$ \it{Department of Mathematical Physics, LTH, 
Lund University, S-22100, Lund, Sweden}} 
\centerline{${}^b$ \it{Science Education Laboratory, Ehime University, 
790-8577 Matsuyama, JAPAN}} 
\vskip 1 cm          
\centerline{\bf ABSTRACT}\par       
\vskip 0.5 cm   
We investigate neutrino masses and mixings within the framework of 
the Zee mass matrix, with three lepton flavors. It is shown that the 
bi-maximal solution is the only possibility to reconcile 
atmospheric and solar neutrino data, within this ansatz. 
We obtain two almost degenerate neutrinos, which are mixtures of all three 
neutrino flavors, with heavy masses 
$\simeq \sqrt{\Delta m_{atm}^2}$. The predicted mass of the lightest 
neutrino, which should consist mostly of $\nu_{\mu}$ and $\nu_{\tau}$, is 
$\simeq \Delta m_{\odot}^2/(2\sqrt{\Delta m_{atm}^2})$. 
\\ \\ \noindent
{\it PACS:} 14.60.Pq, 14.60.Lm, 12.60.Fr \\ \noindent
\newpage 
\topskip 0. cm 
\voffset = -.30 in 
\hoffset = -.10 in 
In the past few years stronger experimental signals, than ever before, have 
been seen for neutrino 
oscillations. Recent atmospheric neutrino experiments~\cite{SKAM1,MACRO} 
indicate oscillations among neutrino flavors with large mixing 
angles~\cite{SKAM2}. 
The simplest solution to the solar neutrino deficit problem, observed by
the Super-Kamiokande experiment ~\cite{SKAM3} as well as 
other experiments~\cite{Solar}, is again neutrino oscillations. 
Indeed, the field of neutrino oscillations is expected to enter into a new
era, with the start of long baseline (LBL) neutrino 
experiments~\cite{K2K,MINOS,ICARUS}. These experiments will hopefully 
solve the present neutrino anomalies. 
\par 
Since the CHOOZ experiment~\cite{CHOOZNew} excludes oscillation of 
$\nu_\mu \ra \nu_e$ with a large mixing angle for 
$\Delta m^2 \geq 9\times 10^{-4} \eV^2$, 
large mixing between $\nu_\m$ and $\nu_\tau$ is the simplest 
interpretation of the atmospheric $\nu_\mu$ deficit. 
The difference between the quark-mixing and lepton-mixing matrices is 
striking. The Cabibbo-Kobayashi-Maskawa mixing matrix $V_{CKM}$\cite{CKM} 
in the quark sector is described by small mixing angles among different 
flavors but in the leptonic sector~\cite{MNS} at least one large 
mixing angle seems to be needed. It will be important to understand why 
these patterns are so different. 
\par
Another issue, to be understood, is why the neutrino masses are so small? 
The most popular answer to the latter question is given by the see-saw 
mechanism~\cite{see-saw} which introduces heavy right-handed Majorana 
neutrinos with masses of the order $10^{10}-10^{16}$GeV. This attractive 
model has been extensively studied in the literature. 
However, it is important to consider also other possible scenarios with 
small neutrino masses, specially extensions of the standard model (SM) at 
a low energy scale. The Zee model~\cite{Zee} is such an alternative 
and has been studied in the literatures for almost twenty 
years~\cite{Wol,Petcov,Pal,Tao,Smirnov}. 
In this paper, we will discuss the present status of the Zee mass matrix, 
in the light of recent experimental results. 
\par 
In the Zee model~\cite{Zee} neutrino masses are generated by radiative 
corrections, and hence the model may provide an explanation of the 
smallness of neutrino masses. In this model, 
the following Lagrangian is added to the SM; 
\bea 
{\cal{L}} &=& \sum_{l,l'=e,\mu,\tau}f_{ll'}\bar{\Psi_{lL}}i\sigma_2 
(\Psi_{l'L})^ch^- + \mu \Phi_1^Ti\sigma_2\Phi_2h^- +h.c. \nonumber\\ 
&=& 2f_{e\mu}[{\bar{\nu_{eL}}}(\mu_L)^c-{\bar e_L}(\nu_{\mu L})^c]h^- 
+ 2f_{e\tau}[{\bar{\nu_{eL}}}(\tau_L)^c-{\bar e_L}(\nu_{\tau L})^c]h^- 
\nonumber\\ 
& & + 2f_{\mu\tau}[{\bar{\nu_{\mu L}}}(\tau_L)^c 
-{\bar \mu_L}(\nu_{\tau L})^c]h^- 
+ \mu(\Phi_1^+\Phi_2^0-\Phi_1^0\Phi_2^+)h^- +h.c. \ , 
\label{lag} 
\eea 
where $\Psi_{lL}=(\nu_l, l)_L^T$, $\Phi_i=(\Phi_i^+, \Phi_i^0)^T$, $i=1,2$.
The Higgs potential is omitted here.   
The charged Zee boson, $h^\pm$, is a singlet under $SU(2)_L$.   
We need at least two Higgs doublets in order to make the Zee 
mechanism viable, since the antisymmetric coupling to the Zee boson is the 
cause of $B-L$ violation, and hence of Majorana masses. Note that only 
$\Phi_1$ couples to leptons, as in the SM. 
The mass matrix, generated by radiative correction at one loop level 
~\cite{Zee,Wol,Petcov,Pal,Tao,Smirnov}, is given by 
\be 
\left(\matrix{0 & m_{e\mu} & m_{e\tau} \cr m_{e\mu} & 0 & m_{\mu\tau} \cr
m_{e\tau} & 
m_{\mu\tau} & 0 \cr}\right) \:\:, 
\label{ZeeMass1} 
\ee 
where 
\bea
m_{e\mu}&=&f_{e\mu}(m_\mu^2-m_e^2)\frac{\mu v_2}{v_1}F(M_1^2,M_2^2) \nonumber\\
m_{e\tau}&=&f_{e\tau}(m_\tau^2-m_e^2)\frac{\mu v_2}{v_1}F(M_1^2,M_2^2) 
\label{Zeemass2} \\
m_{\mu\tau}&=&f_{\mu\tau}(m_\tau^2-m_\mu^2)\frac{\mu v_2}{v_1}F(M_1^2,M_2^2) 
\nonumber
\eea
and
\be
F(M_1^2,M_2^2)=\frac{1}{16\pi^2}\frac{1}{M_1^2-M_2^2}ln\frac{M_1^2}{M_2^2}.
\nonumber 
\ee 
The parameter $v_{1(2)}$ is the vacuum expectation value of the neutral 
component of the Higgs doublet $\Phi_{1(2)}$. $M_1$ and $M_2$ are 
the masses of the physical particles defined by the fields 
\be 
H_1^+=h^+\cos\phi-\Phi^+\sin\phi \;\;\ , \;\;\;\;\; 
H_2^+=h^+\sin\phi+\Phi^+\cos\phi \;, 
\ee 
where $\Phi^+$ is the charged Higgs boson that would have been a physical 
particle in the absence of the $h^+$. 
Finally, the mixing angle $\phi$ is defined by 
\be 
\tan2\phi=\frac{4\sqrt{2}\mu M_W}{g\sqrt{(M_1^2-M_2^2)^2- 
(4\sqrt{2}g^{-1}\mu M_W)^2}} \;. 
\ee 
Due to the antisymmetry of the coupling matrix, $f_{ll'}=-f_{l'l}$, the 
Zee model requires all diagonal elements in Eq. (\ref{ZeeMass1}) to 
vanish at one loop level. Small corrections will however be obtained at 
higher orders in perturbation theory. Hereafter, we refer to the above 
matrix, with vanishing diagonal elements, as the Zee mass matrix. 
\par 
The parameters $m_{e\mu}, m_{e\tau}, m_{\mu\tau}$ in the Zee mass matrix 
are not described by the eigenvalues $m_i\; (i=1,2,3)$ 
due to the traceless property of the matrix, leaving only two independent 
observable parameters. 
In the same way it is impossible to represent the mixing matrix $U$ 
by using $m_i$. 
In literature~\cite{Wol,Petcov,Pal,Tao,Smirnov}, 
it has been assumed that there is a hierarchy such as 
$m_{e\mu}\ll m_{e\tau}, m_{\mu\tau}$ 
and the neutrino masses and mixings are discussed under such assumptions. 
This hierarchy is natural if the coupling constants 
$f_{ll'}$ are of the same order of magnitude. 
However, we would like to explore all possibilities of masses and mixings 
in the Zee model by relaxing this assumption. 
Instead we use recent atmospheric neutrino data~\cite{SKAM1,MACRO} 
as input to determine patterns in the Zee mass matrix that are viable. 
\par 
The neutrino mass matrix $M_{\nu}$ is generally, for Majorana particles, 
constructed by 
\be 
M_{\nu}=UM^{diag.}U^T
\label{MLL}
\ee
due to its symmetric property $M_\nu=M_\nu^T$. The mixing matrix $U$, 
called the MNS mixing matrix~\cite{MNS}, is defined in the basis where 
the mass matrix of charged leptons is diagonal, with masses 
$m_{e,\mu,\tau}$. Furthermore, 
\be 
M^{\rm diag.}=\left(\matrix{m_1 & 0 & 0 \cr 
0 & m_2 & 0 \cr 
0 & 0 & m_3 \cr}\right) \ . 
\ee 
For Majorana neutrinos there are three phases in the matrix $U$ and 
this is generally given by 
\be 
U = 
\left(\matrix{c_1c_3 & s_1c_3e^{i\b}  & s_3e^{-i(\d-\a)} \cr
      -s_1c_2-c_1s_2s_3e^{i\d} & (c_1c_2-s_1s_2s_3e^{i\d})e^{i\b} 
& s_2c_3e^{i\a} \cr
       s_1s_2-c_1c_2s_3e^{i\d} & (-c_1s_2-s_1c_2s_3e^{i\d})e^{i\b} 
& c_2c_3e^{i\a} \cr}
\right) \; , \label{U}
\ee
where $c_i \equiv \cos\th_i$ and $s_i \equiv \sin\th_i$.
The Zee mass matrix exhibits no $CP$ violation~\cite{Bilenky} and we will 
therefore neglect the phases in our investigation. 
The diagonal elements in $M_{\nu}$ are given by 
\bea\label{mass1} 
(1,1)\:\:& &~ m_1c_1^2c_3^2+m_2s_1^2c_3^2+m_3s_3^2 \;, \nonumber\\ 
(2,2)\:\:& &~ m_1(s_1c_2+c_1s_2s_3)^2+m_2(c_1c_2-s_1s_2s_3)^2+m_3s_2^2c_3^2
\;, \\ 
(3,3)\:\:& &~ m_1(s_1s_2-c_1c_2s_3)^2+m_2(c_1s_2+s_1c_2s_3)^2+m_3c_2^2c_3^2
\;. 
\nonumber 
\eea 
These should be zero (in general small) in the Zee model and we 
arrive at the following relations:
\be
m_2=-\frac{\cos^2\th_1-\tan^2\th_3}{\sin^2\th_1-\tan^2\th_3}m_1
\;, \qquad 
m_3=-m_1-m_2 \;.
\label{massrelation}
\ee
The second equality is obvious by the traceless property of the 
\begin{figure}[tb] 
\begin{center} 
\input{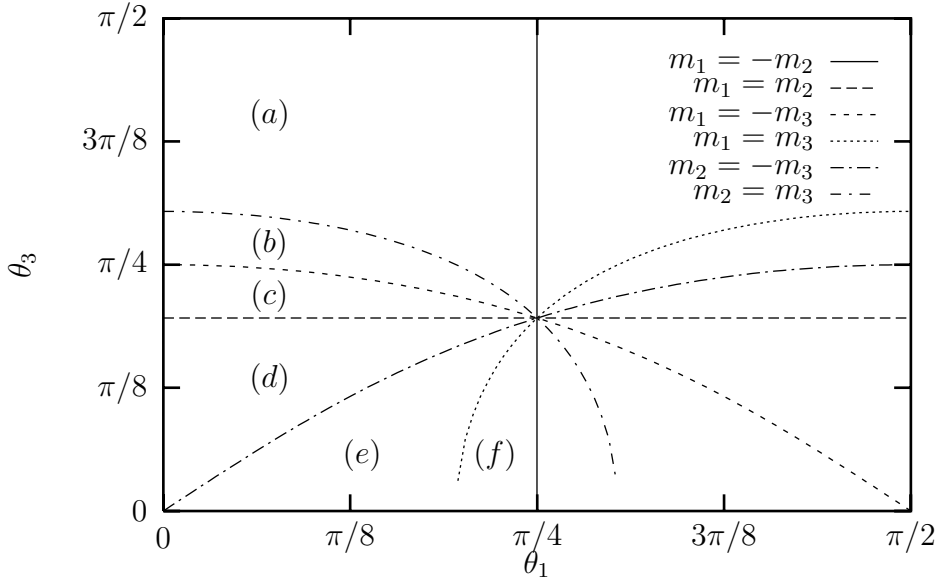} 
\end{center} 
\caption{The dependence of mass eigenvalues on the mixing angles $\th_1$ 
and $\th_3$. The lines represents special relations among the three masses 
as denoted in the figure. In the right hand side the domains are running 
oppositely, starting with (a) in the bottom and ending with (f) in the top.
In region (a) we have $m_1>-m_2>-m_3$, (b) $m_1>-m_3>-m_2$, 
(c) $-m_3>m_1>m_2$, (d) $-m_3>m_2>m_1$, (e) $-m_2>m_3>m_1$, 
and (f) $-m_2>m_1>m_3$. Here the sign of $m_1$ is taken to be positive.
\label{fig1}}
\end{figure} 
Zee mass matrix. Two of three masses have the same sign and the remaining 
one has opposite sign, which implies that one of the fields has 
opposite CP parity as compared to the other two. 
The dependence of mass eigenvalues on $\th_1$ and $\th_3$ is shown 
in Fig.\ref{fig1}. 
Inserting (\ref{massrelation}) into (\ref{mass1}) gives the relation 
\be 
\cos2\th_1\cos2\th_2\cos2\th_3=
\frac{1}{2}\sin2\th_1\sin2\th_2(3\cos^2\th_3-2)\sin\th_3 . 
\label{angle} 
\ee 
This equation means that the three mixing angles are not independent. 
For typical values of $\th_1, \th_2, \th_3$ the structure of 
the mixing matrix is discussed later, using this equation. 
\vskip 0.3cm 
\par 
Before entering into the analysis of the Zee mass matrix, we give a short 
survey of recent neutrino experiments. Our approach is to assume that 
oscillations account for the solar and atmospheric neutrino data, thus 
pinning down two mass squared differences, which is the maximal number 
of mass differences in the model we are investigating. If the results of 
LSND~\cite{LSND} would be confirmed by KARMEN~\cite{KARMEN} or any other 
experiment, the model examined in this paper would no longer be relevant. 
The deficit of $\nu_\mu$ in recent Super Kamiokande data \cite{SKAM1}, 
is interpreted as oscillation of $\nu_\mu \rightarrow \nu_\tau$ 
with nearly maximal mixing angle in a two flavor analysis 
\cite{SKAM2,Fogli}. 
These results yield 
\be 
\D m^2_{atm}\simeq (0.5-6)\times 10^{-3}\eV^2 \;,\qquad \sin^22\th_{atm}>0.82 
\quad (90\% C.L) \;. 
\label{atm} 
\ee 
Further, the deficit in the solar neutrino experiments suggests the 
following best-fit solutions \cite{Bahcall} as \\ \\ 
\noindent
(1) MSW small angle solution; 
$\D m^2_{\odot} \simeq 5.4 \times 10^{-6}\eV^2 \:, 
 \sin^22\th_{\odot}\simeq 6\times 10^{-3} \:, $ \\
\noindent
(2) MSW large angle solution;
$\D m^2_{\odot}\simeq 1.8 \times 10^{-5}\eV^2 \:, 
\sin^22\th_{\odot}\simeq 0.76 \ ,$ \\
\noindent
(3) ''just-so'' vacuum solution;
$ \D m^2_{\odot}\simeq 6.5 \times 10^{-11}\eV^2 \:, 
\sin^22\th_{\odot}\simeq 0.75 \ .$ \\ \\ \noindent 
It is noted that the large angle MSW solution seems to be excluded by the 
simultaneous fits to all the available data~\cite{Bahcall}.
As there are still theoretical uncertainties about this~\cite{Bahcall2}, we 
will keep this possibility in our considerations.
The combined results of atmospheric and solar neutrino experiments suggest 
that there exist two hierarchical mass squared differences 
$\D m^2_{atm}\gg \D m^2_\odot$. 
Further, the component $U_{e3}$ in Eq.(\ref{U}) should be small, 
as suggested by the CHOOZ experiment~\cite{CHOOZNew}. 
The value $\sin^22\th_{CHOOZ}<0.18$ implies $|U_{e3}|<0.22$ for 
$\D m^2 \ge 9 \times 10^{-4}{\rm eV}^2$~\cite{Giunti1}. 
For the MSW small angle solution, case (1), together with the atmospheric 
results, a possible mixing matrix has the form 
\begin{equation} 
U_1\simeq \left (\matrix{1 & \e_1 & \e_2 \cr 
\e_3 & c & s \cr 
\e_4 & -s & c \cr} \right ) \ , 
\label{U1} 
\end{equation} 
where $c \simeq s \simeq 1/\sqrt{2}$. 
This mixing matrix might be realized in the case; 
\be 
|\Delta m_{32}^2|\simeq|\Delta m_{31}^2|\simeq \Delta m^2_{atm}\;,
 \qquad 
|\Delta m_{21}^2|\simeq \Delta m^2_{\odot} \ , 
\ee 
which demands 
$m_3 \gg m_i \geq m_j$ or $m_i \geq m_j \gg m_3 \; (i,j=1,2$ or $2,1)$ 
or 
\be 
|\Delta m_{32}^2|\simeq|\Delta m_{21}^2|\simeq \Delta m^2_{atm}\;,
 \qquad 
|\Delta m_{31}^2|\simeq \Delta m^2_{\odot} \ , 
\ee 
implying $m_2 \gg m_i \geq m_j$ or $m_i \geq m_j \gg m_2\;(i,j=1,3$ 
or $3,1)$. Another solution for MSW small angle mixing is 
\begin{equation} 
U_1^{'}\simeq \left (\matrix{ \e_1 & 1 & \e_2 \cr 
c & \e_3 & s \cr 
-s & \e_4 & c \cr} \right ) \ . 
\label{U1p} 
\end{equation} 
This type of mixing suggests 
\be 
|\Delta m_{32}^2|\simeq|\Delta m_{31}^2|\simeq \Delta m^2_{atm}\;,
 \qquad 
|\Delta m_{21}^2| \simeq \Delta m^2_{\odot} 
\ee 
or 
\be 
|\Delta m_{31}^2|\simeq|\Delta m_{21}^2|\simeq 
\Delta m^2_{atm}\;, \qquad |\Delta m_{32}^2| \simeq \Delta m^2_{\odot} 
\ee 
and we obtain the solution 
$m_1 \gg m_i \geq m_j$ or $m_i \geq m_j \gg m_1 \; (i,j=2,3$ or $3,2)$ 
for the latter case. 
\par 
For the ``just-so'' and MSW large angle solutions a typical mixing matrix 
is 
\begin{equation} 
\left (\matrix{ c & s & \e_1 \cr -s & c & \e_2 \cr \e_3 & \e_4 & 1 \cr} 
\right ) 
\quad {\rm or} \quad 
\left (\matrix{ c & s & \e_1 \cr \e_2 & \e_3 & 1 \cr -s & c & \e_4 \cr} 
\right ).  
\end{equation}   
Neither of these is compatible with the maximal mixing pattern 
of the atmospheric $\nu_\mu \rightarrow \nu_\tau$ oscillation, and must be 
discarded for this reason. 
We are left with one possibility;
\be
U_3\simeq
\left (\matrix{ c' & s' & \e \cr  -cs' & cc' & s \cr 
ss' & -sc' & c \cr} \right )   
\label{U3}
\ee
for interpreting large angle solar neutrino solutions, where 
$c\simeq s \simeq c'\simeq s'\simeq 1/\sqrt{2}$. In the limit $\epsilon = 0$ 
this is known as the bi-maximal mixing matrix~\cite{Bimax}. Nearly 
bi-maximal mixing is discussed in Ref.\cite{NearBimax}.
Taking 
\be 
|\Delta m_{32}^2|\simeq|\Delta m_{31}^2|\simeq \Delta m^2_{atm}, 
\quad |\Delta m_{21}^2| \simeq \Delta m^2_{\odot} 
\ee 
with $ m_3 \gg m_i \geq m_j$ or $m_i \geq m_j 
\gg m_3,\;(i,j=1,2\; {\rm or}\; 2,1)$, 
it has been shown that the mixing matrix $U_3$ is consistent with 
vacuum solution for both solar and atmospheric neutrino anomalies within 
the experimental uncertainties ~\cite{Giunti}. The MSW large angle 
solution could also be accommodated here, but then the allowed parameter 
space is rather small. 
\par 
Furthermore the degenerate case $m_1 \simeq m_2 \simeq m_3$ is another 
possibility for all the above cases. Using the definition; 
\be 
m_a=m_0\;, \quad m_b=m_0(1-\frac{\d}{2})\;, \quad
 m_c=m_0(1-\frac{\d+\e}{2}) , 
\label{DGN} 
\ee 
we can assign $a,b,c$ to each of $1,2,3$ according to the mass relation 
obtained above.
Only in this degenerate case is it possible to have a mass 
$m_0={\cal{O}}(1)$eV for the neutrinos, as is suggested by the hot dark 
matter argument. Setting $m_0=1$eV the parameters $\d$ and $\e$ take the 
values ${\cal O}(10^{-3}-10^{-2})$ and ${\cal O}(10^{-5})$ respectively 
to reproduce $\D m^2_{ab}\simeq  \D m^2_{ac}\simeq \D m^2_{atm}$ and 
$\D m^2_{bc}\simeq \D m^2_\odot$.      
\vskip 0.3cm
\par
We now return to the analysis of the Zee mass matrix.
In accordance with the data of atmospheric neutrino experiments 
we require $\th_2\simeq\pi/4$. 
Now Eq.(\ref{angle}) implies that we have the following possibilities; 
$\th_3\simeq 0$ or $\th_1\simeq 0$ or $\th_3\simeq \arctan\sqrt{1/2}$.  
The latter case will be commented on later. We will only consider the case of 
$\th_1\simeq 0$ together with $\th_3\simeq 0$, due to the 
experimental constraints on $U_{e3}$. 
To begin with we concentrate on the solution $\th_3\simeq 0$. 
\par 
If we take the extreme limit $\th_2=\pi/4$ and $\th_3=0$, the mixing matrix
becomes 
\be 
U_{atm}\simeq \left(\matrix{c_1 & s_1 & 0 \cr 
-\frac{s_1}{\sqrt{2}} 
& \frac{c_1}{\sqrt{2}} & 
\frac{1}{\sqrt{2}} \cr
               \frac{s_1}{\sqrt{2}} 
& -\frac{c_1}{\sqrt{2}} & \frac{1}
{\sqrt{2}} \cr}\right).
\label{Uatm}
\ee
It is noted that the extreme case does not change the qualitative 
structure of the model and is consistent with combined SK and CHOOZ  
data in a three flavor analyses~\cite{SKAM2,Fogli}.    
In this limit we obtain the constraints 
\be
m_1c_1^2+m_2s_1^2=0 \:,\qquad m_1s_1^2+m_2c_1^2+m_3=0
\label{Con1}
\ee
from Eq.(\ref{mass1}). Then the parameters satisfy the constraint
\be
\tan^2\th_1=-\frac{m_1}{m_2}>0 \;,
\label{mass3}
\ee
and we arrive at
\bea
M_{\nu}&=&\left(\matrix{
0 & -\frac{c_1s_1}{\sqrt{2}}(m_1-m_2) &\frac{c_1s_1}{\sqrt{2}}(m_1-m_2) \cr
-\frac{c_1s_1}{\sqrt{2}}(m_1-m_2) & 0 & -\frac{1}{2}(m_1s_1^2+m_2c_1^2-m_3) \cr
\frac{c_1s_1}{\sqrt{2}}(m_1-m_2) & -\frac{1}{2}(m_1s_1^2+m_2c_1^2-m_3) & 0
\cr}\right) \nonumber\\
&=&\left(\matrix{
0 & \pm\sqrt{\frac{|m_1m_2|}{2}} & \mp\sqrt{\frac{|m_1m_2|}{2}} \cr
\pm\sqrt{\frac{|m_1m_2|}{2}} & 0 & -m_1-m_2 \cr
\mp\sqrt{\frac{|m_1m_2|}{2}} & -m_1-m_2 & 0 \cr}\right) ,
\label{MLLZee}
\eea
where the upper (lower) sign corresponds to the case 
$m_1<0$ $(m_1>0)$.
The mixing matrix becomes
\be
U=\left(\matrix{
   \sqrt{\frac{|m_2|}{|m_1|+|m_2|}} & \sqrt{\frac{|m_1|}{|m_1|+|m_2|}} & 0 \cr
  -\sqrt{\frac{|m_1|}{2(|m_1|+|m_2|)}} & \sqrt{\frac{|m_2|}{2(|m_1|+|m_2|)}} & 
  \frac{1}{\sqrt{2}} \cr
   \sqrt{\frac{|m_1|}{2(|m_1|+|m_2|)}} & -\sqrt{\frac{|m_2|}{2(|m_1|+|m_2|)}} &
  \frac{1}{\sqrt{2}} \cr}\right) .
\ee
\par 
As the two mass squared differences are hierarchical we have to stay 
close to the lines in  Fig.(\ref{fig1}). 
Due to the symmetry we only have to survey the left part 
of the parameter region, i.e $\th_1 \in [0,\pi/4]$. Changing from the 
left to right side merely corresponds to an interchange of 
$m_1 \leftrightarrow m_2$ and has no physical effect. 
\newcounter{case}
\setcounter{case}{1}
We now consider the three possible values for $\th_1$. \\ \noindent
(\Roman{case}).~ \stepcounter{case} 
The case with $\th_1\simeq 0$. The Zee mass matrix 
then requires the following mass relations to hold;
\be
|m_2| \simeq |m_3| \gg |m_1| \quad , \:\:
\D m^2_{21}\simeq \D m^2_{31}=\D m^2_{atm}\ ,\quad
\D m^2_{23}=\D m^2_{\odot} \label{relation1} \ .
\ee
In this case we get the mixing matrix;
\be
U_{1}^{Zee}=\left(\matrix{1 & \e_1 & 0 \cr
-\frac{\e_1}{\sqrt{2}} & \frac{1}{\sqrt{2}} & \frac{1}{\sqrt{2}} \cr
\frac{\e_1}{\sqrt{2}} & -\frac{1}{\sqrt{2}} & \frac{1}{\sqrt{2}} 
\cr}\right) ,
\label{UZee1} 
\ee
with $\e_1\simeq \sqrt{|m_1/m_2|}$.
This corresponds to the small angle MSW solution $U_1$ (Eq.(\ref{U1})).
Nevertheless due to the mass relations in Eq.(\ref{relation1}) the predicted 
probability of $\nu_\mu \ra \nu_\tau$ oscillation,
\bea
P(\nu_\mu \ra \nu_\tau)&=& -4\sum_{i>j} U_{\mu i}U_{\tau j}^*
U_{\mu j}^*U_{\tau i}\sin^2\frac{\Delta m_{ij}^2L}{4E} \nonumber \\
&\simeq&4U^2_{\mu 1}U^2_{\tau 1}\sin^2\frac{\D m^2_{atm}L}{4E} \\
&=&\e_1^4\sin^2\frac{\D m^2_{atm}L}{4E} \nonumber \;, 
\eea
is tiny in contradiction with the SK experiment. Here and in the following 
we neglect terms with $\D m_{\odot}^2$ in $P(\nu_\mu \ra \nu_\tau)$. 
Thus, the Zee mass matrix and large angle solution are not compatible, 
for $\th_1 \simeq 0$. 
Setting $\th_1\simeq \pi/2$ would correspond to 
$m_1 \leftrightarrow m_2$ compared to the current case. 
It also causes an interchange of the two first columns in $U_1^{Zee}$, 
which yields the matrix in Eq.(\ref{U1p}). However this must be discarded 
for the same reason. 
\\ \noindent
(\Roman{case}).~ \stepcounter{case}
Here we take $\th_1\simeq \arctan 1/\sqrt{2}$ and obtain 
\be
|m_1| \simeq |m_3| \gg |m_2| \quad , \:\:
\D m^2_{12}\simeq \D m^2_{32}=\D m^2_{atm}\ ,\quad 
\D m^2_{13}=\D m^2_{\odot}\ .
\ee
This case requires the mixing matrix to be 
\be
U_2^{Zee}=\left(\matrix{\sqrt{\frac{2}{3}} & \sqrt{\frac{1}{3}} & 0 \cr
-\sqrt{\frac{1}{6}} & \sqrt{\frac{1}{3}} &  \sqrt{\frac{1}{2}} \cr
\sqrt{\frac{1}{6}} & -\sqrt{\frac{1}{3}}&
\sqrt{\frac{1}{2}} \cr}\right) .
\label{UZee2}
\ee
Giving the angles:
\be
 \sin^22\th_{\rm{CHOOZ}} = 4U_{e2}^2(1-U_{e2}^2) \simeq \frac{8}{9} \;,
\qquad 
 \sin^22\th_{atm} = 4U_{\mu 2}^2U_{\tau 2}^2 \simeq \frac{4}{9} \;.
\ee
Hence this is also incompatible with experiments.
\\ \noindent
(\Roman{case}).~ \stepcounter{case}
In this third case $\th_1 \simeq \pi/4$, whereby
\be
|m_1| \simeq |m_2| \gg |m_3| \quad , \:\:
\D m^2_{13}\simeq \D m^2_{23}=\D m^2_{atm}\ ,\quad 
\D m^2_{12}=\D m^2_{\odot}\ .
\ee
The mixing matrix is
\be
U_3^{Zee}=\left(\matrix{\frac{1}{\sqrt{2}} & \frac{1}{\sqrt{2}} & 0 \cr
-\frac{1}{2} & \frac{1}{2} & \frac{1}{\sqrt{2}} \cr
\frac{1}{2} & -\frac{1}{2} &
\frac{1}{\sqrt{2}} \cr}\right) .
\label{UZee3}
\ee
This corresponds to the solution given by $U_3$ in Eq.(\ref{U3}).
The oscillation probability reads
\be
P(\nu_\mu \ra \nu_\tau) \simeq \sin^2\frac{\D m^2_{atm}L}{4E} \;,  
\ee
in good agreement with experiments. 
Therefore we have a unique solution compatible with large angle in 
$\nu_\mu \ra \nu_\tau$ within the ansatz of the Zee mass matrix.
Due to the traceless property this implies
\be 
|m_1|\simeq |m_2|\simeq \sqrt{\D m_{atm}^2}, \quad 
|m_3|\simeq \frac{\D m^2_\odot}{2\sqrt{\D m_{atm}^2}} \;,
\label{numerical}
\ee
This is the case of ``pseudo-Dirac'' since $m_1\simeq -m_2$.
The probabilities for other oscillation processes are as follows: 
\bea
P(\nu_e \ra \nu_e) &\simeq& 1-\sin^2\frac{\D m^2_{\odot}L}{4E} \nonumber\\
P(\nu_e \ra \nu_\mu) &\simeq& P(\nu_e \ra \nu_\tau) 
\simeq \frac{1}{2}\sin^2\frac{\D m^2_{\odot}L}{4E} \;.
\eea
Thus the solar neutrinos are 
converted into an equal amounts of $\nu_\mu$ and $\nu_\tau$.
\par
The mixing patterns $U_{1}^{Zee}$ corresponds to the small angle MSW 
solution, and the matrix $U_3^{Zee}$ corresponds to the large angle MSW  
or "just-so" solution. It is interesting to notice that the solution of 
three degenerate masses with ${\cal{O}}(1)\eV$ is not allowed due to the 
relation $m_3=-m_1-m_2$ and Eq.(\ref{mass3}).
This model requires naturally a hierarchical structure for mass matrix 
in the case of $\th_2=\pi/4$ and $\th_3=0$.
In conclusion only the bi-maximal solution, given in Eq.(\ref{UZee3}), is 
feasible, within the framework of the Zee mass matrix, whereas the 
solutions like $U_{1}^{Zee}$ and $U_{2}^{Zee}$ are not allowed.
\par
\vskip 0.3cm
Above, we have analyzed the Zee mass matrix at the limit $\th_3=0$. 
It is also important to discuss the case of nonzero $\th_3$ in order to 
obtain three flavor angles and to have predictions for future experiments. 
We parameterize this as
\be
\th_1=\frac{\pi}{4}+\d_1 \;,\qquad \th_2=\frac{\pi}{4}+\d_2\;,\qquad \th_3=\d_3
\ee
with $\d_i\;(i=1,2,3) \ll 1$ and we will neglect terms of order $\d^2_i$. 
The traceless property of the Zee mass matrix requires the relation
$\d_3\simeq 8\d_1\d_2$, by using Eq.(\ref{angle}), and yields the
approximate mixing matrix 
\be
U_{Zee}\simeq\left(\matrix{\frac{1-\d_1}{\sqrt{2}}c_3
& \frac{1+\d_1}{\sqrt{2}}c_3 & \d_3 \cr
-\frac{1+\d_1-\d_2+\d_3}{2} & \frac{1-\d_1-\d_2-\d_3}{2} 
& \frac{1+\d_2}{\sqrt{2}}c_3 \cr
\frac{1+\d_1+\d_2-\d_3}{2} & -\frac{1-\d_1+\d_2+\d_3}{2} & 
\frac{1-\d_2}{\sqrt{2}}c_3 \cr}
\right) \ ,
\label{UZee}
\ee
where $c_3=\sqrt{1-\d_3^2}$ and the mass matrix is described as
\be
M_{\nu}\simeq\left(\matrix{0 & -\frac{1-2\d_1-\d_2}{\sqrt{2}} & 
 \frac{1-2\d_1+\d_2}{\sqrt{2}}\cr
-\frac{1-2\d_1-\d_2}{\sqrt{2}} & 0 & -4\d_1 \cr
\frac{1-2\d_1+\d_2}{\sqrt{2}} & -4\d_1 & 0 \cr}\right)m_1 \ .
\label{MLLZee2}
\ee
The ratio of lightest mass to heavy mass is approximately 
given by $\d_1$ as shown in Eqs.(\ref{MLLZee},\ref{MLLZee2}). 
Which restrictions do the present experiments give for the parameters 
$\d_1, \d_2, \d_3$? 
Taking Eq.(\ref{atm}) and the lower limit for large angle solution of solar
neutrino as $\sin^22\th_\odot>0.65$ we obtain the following constraints;
\bea
\sin^22\th_{atm}&\simeq& 4|U_{\mu 3}|^2|U_{\tau 3}|^2\simeq \frac{(1-2\d_2^2)
(1-2\d_3^2)}
{2}\ge 0.82 \;, \nonumber\\
\sin^22\th_\odot&\simeq& 4|U_{e1}|^2|U_{e2}|^2\simeq \frac{(1-2\d_1^2)
(1-2\d_3^2)}{2}\ge 0.65 \ .
\eea
These inequalities with $|\d_3|\simeq 8|\d_1\d_2|$ leads to
\be
\d_{1(2)}^2\leq \frac{1}{128\d_{2(1)}^2}\left(1-\frac{0.82(0.65)}
{1-2\d_{2(1)}^2}\right) \ .
\label{limit}
\ee
We obtain the upper limit for $\d_3$ to be 
\be
|\d_3|\leq 0.28 \ .
\ee
If the data will be improved as $\sin^22\th_{atm}>0.95,
\;\sin^22\th_{\odot}>0.95$ we get the upper limit $|\d_3|<0.1$. 
A better estimate can nevertheless be deduced by noticing the following. 
The eigenvalues of $M_{\nu}$ are $m_1, -(1-4\d_1)m_1, 4\d_1m_1$. 
In order to adjust the mass squared difference of $\D m^2_\odot$ using 
Eq(\ref{numerical}), $\d_1$ should be chosen as ${\cal{O}}(10^{-3})$ for 
MSW or ${\cal{O}}(10^{-8})$ for ``just-so''. 
This demands also $\d_3$ to be tiny due to the relation 
$\d_3\simeq 8\d_1\d_2$.
The value $\d_3\simeq {\cal{O}}(10^{-3})$ or ${\cal{O}}(10^{-8})$ is well  
within the present experimental upper limit. In this case we expect 
no $\nu_\mu \ra \nu_e$ oscillations at the LBL 
experiments~\cite{K2K,MINOS,ICARUS} since,   
\be
P(\nu_\mu \ra \nu_e)\simeq 2\d_3^2\sin^2\frac{\D m^2_{atm}L}{4E}\simeq 0 \ .
\ee
\vskip 0.3cm
\par
We now briefly discuss the remaining solution: 
$\th_3=\arctan (1/\sqrt{2})+\d_3,\; 
m_1\simeq m_2 \simeq -m_3/2$, when requiring $\th_2=\pi/4+\d_2$ in 
the Zee mass matrix. Here we take $\th_1=\pi/4+\d_1$ to obtain the ``maximal'' 
case~\cite{Maximal}. 
The mixing matrix reads
\be
U\simeq\left(\matrix{\frac{1-\d_1}{\sqrt{3}} & \frac{1+\d_1}{\sqrt{3}} & \frac{1}{\sqrt{3}} \cr
-\frac{1+\d_1-\d_2}{2}-\frac{1-\d_1+\d_2}{2\sqrt{3}} &
\frac{1-\d_1-\d_2}{2}-\frac{1+\d_1+\d_2}{2\sqrt{3}}  &  
\frac{1+\d_2}{\sqrt{3}} \cr
\frac{1+\d_1+\d_2}{2}-\frac{1-\d_1-\d_2}{2\sqrt{3}} &
-\frac{1-\d_1+\d_2}{2}-\frac{1+\d_1-\d_2}{2\sqrt{3}} &  
\frac{1-\d_2}{\sqrt{3}} \cr}\right)
\ee
and
\be
M_\nu\simeq-\left(\matrix{
0 & 1-(2\sqrt{3}-1)\d_2 & 1-(2\sqrt{3}+1)\d_2 \cr
1-(2\sqrt{3}-1)\d_2 &  0 & 1-2\sqrt{3}\d_2 \cr
1-(2\sqrt{3}+1)\d_2 & 1-2\sqrt{3}\d_2 &  0 \cr}\right)m_1 \ .
\ee
Here $\d_3\simeq -2\sqrt{2}\d_1\d_2/\sqrt{3}$, from Eq.(\ref{angle}), and therefore 
terms with $\d_3$ are neglected in the above matrices. 
This ``maximal'' case is only allowed for $\D m^2<9\times 10^{-4}\eV^2$ 
due to results of the CHOOZ experiment~\cite{CHOOZNew}. 
However, a three flavor analysis~\cite{Fogli} without CHOOZ data 
shows that the ``maximal'' solution lies in the region 
\be 
1.5 \times 10^{-3} \leq \D m^2_{32}\simeq \D m^2_{31} \simeq 3m_1^2 \leq 
6.5 \times 10^{-3}\eV^2 
\ee 
at 90\%C.L. 
Combining results from CHOOZ and SK experiments tends to exclude this 
``maximal'' solution, which predicts large effects for 
$P(\nu_\mu \ra \nu_e)$ in LBL experiments. 
\vskip 0.3cm 
\par 
A comment on neutrino-less double beta decay is now in order. 
Although we get the heaviest mass 
$|m_1|\simeq |m_2|\simeq 0.02\sim 0.08 {\rm eV}$ for bi-maximal mixing, 
neutrino-less double beta decay is forbidden in the Zee model. 
The effective neutrino mass in the decay 
\begin{equation} 
\langle m_\nu \rangle = | \sum_i U^2_{ei}m_{\nu i} | 
\end{equation} 
\noindent 
is exactly zero due to the condition that (1,1) entry in the mass matrix 
is zero. Generally, a theory which has zero entry in (1,1) leads to 
vanishing neutrino-less double beta decay, even if the neutrino is a 
Majorana particle~\cite{Moh}. 
\par 
\vskip 0.3cm 
Returning to the Zee model which motivated the use of the 
Zee mass matrix, we find that the requirement of 
nearly bi-maximal mixing necessarily yields 
\be 
M_{\nu}\simeq\left(\matrix{\d_1 & -1   &   1  \cr 
                            -1  & \d_1 & \d_1 \cr  
                             1  & \d_1 & \d_1 \cr}\right)  .
\label{M3}
\ee
By using the relation $m_{e\mu} \simeq m_{e\tau}$, in Eq.(\ref{ZeeMass1}), 
we obtain
\be 
\frac{f_{e\mu}}{f_{e\tau}}\simeq \frac{m_\tau^2}{m_\mu^2} \simeq 2.9 
\times 10^{2}
\label{f1}
\ee
and
\be
\frac{m_{e\tau}}{m_{\mu\tau}}=\frac{f_{e\tau}}{f_{\mu\tau}}
\simeq \frac{\sqrt{2}\D m_{atm}^2}{\D m^2_\odot}\simeq 10^2 \; {\rm or}\; 10^7
\label{f2}
\ee
for $\D m^2_{atm}\simeq 10^{-3}\eV^2$ and 
$\D m^2_\odot\simeq10^{-5}$ or $10^{-10}\eV^2$. 
The magnitude of $f_{e\mu}$ may be estimated using Eq.(3) and 
$0.02<m_{e\mu}<0.08\eV$ 
\be 
f_{e\mu}\simeq (2\sim 7)\left(\frac{100{\rm GeV}}{\mu}\frac{1}{\tan 
\b}\right)\times 10^{-4} 
\ee 
where $\tan\beta=v_2/v_1$ and we have taken $M_1=200{\rm GeV}$ and 
$M_2=300{\rm GeV}$. 
We see that the relation of couplings must be of the form 
\be 
f_{e\mu} \gg f_{e\tau} \gg f_{\mu\tau} 
\label{antih} 
\ee 
in order to agree with experiments. 
This indicates that most likely an extension of the Zee model is needed to
give 
the anti-hierarchy above. This might be realized by assigning an approximate
conserved $U(1)$~\cite{barbieri} charge 
\be
 L\equiv L_e-L_\mu-L_\tau \label{leptonnumber} \;,
\ee
which will strongly suppress $f_{\mu\tau}$. Although this cannot explain the 
factor of $\sim 10^2$ between $f_{e\mu}$ and $f_{e\tau}$, it can account for  
the more demanding factor of $10^7$ in the ``just-so'' case.  
With this extension the matrix in Eq.(\ref{M3}) can be derived from the 
Zee model. The tiny values on the diagonal will be caused by higher 
order effects, as mentioned before. 
Also it is remarkable that this will give an inverse mass hierarchy 
as $|m_3| \ll |m_1| \simeq |m_2|$. 
This pattern is substantially different from those studied in the previous
analyses of the Zee model~\cite{Smirnov,Gaur}. 
In Ref.\cite{Smirnov}, the case $|m_1| \ll |m_2| \simeq |m_3|$ 
is taken and the heavy masses $|m_2|\simeq |m_3|$ 
adjusted to be $(1-5)$eV as the candidates of hot dark matter and 
the mass squared differences are assigned to be 
$\D m^2_{32}=\D m^2_{atm},\:\D m^2_{31}\simeq\D m^2_{31}\simeq \D m^2_{LSND}$. 
The addition of one sterile neutrino to the original Zee model is studied 
in Ref.\cite{Gaur}, giving a possible explanation of 
all positive neutrino oscillation experiments, still keeping the 
assumption of $|f_{e\mu}| \simeq |f_{e\tau}| \simeq |f_{\mu\tau}|$. 
\par
The current constraints on the new parameters 
$f_{e\mu},f_{e\tau},f_{\mu\tau},\mu$ 
in the Zee model are discussed in detail in Ref.\cite{Smirnov}. 
Unfortunately there are not enough experimental data to determine these 
parameters. Here we will only mention about the radiative decays of 
neutrino and charged leptons induced by Zee boson exchange~\cite{Petcov}. 
In the present case the possible radiative decays are 
$\nu_{1(2)}\rightarrow \nu_3+ \gamma$. Under the assumption of 
Eq.(\ref{antih}) the amplitude is given as
\be
A(\nu_1 \ra \nu_3+\gamma)=
\frac{e}{32\pi^2{\bar M^2}}(m_1-m_3)f_{e\mu}^2U_{\mu1}U_{\mu3}
{\bar \nu_3}i\sigma_{\mu\nu}q^\nu\e^\mu\gamma_5\nu_1 ,
\ee
where
\be
\frac{1}{\bar M^2}=\frac{\cos^2\phi}{M_1^2}+\frac{\sin^2\phi}{M_2^2}.
\ee
Here the CP property of $\nu_1$ and $\nu_3$ is taken to be the same.
The decay width and lifetime are 
\bea 
\Gamma(\nu_1 \rightarrow \nu_3+\gamma)&=&\frac{(m_1-m_3)^3}{8\pi}|A|^2
 \nonumber\\ 
&\simeq&\frac{\a}{2}(\frac{f_{e\mu}^2U_{\mu1}U_{\mu3}} 
{32\pi^2{\bar{M^2}}})^2m_1^5 \\ 
\tau(\nu_1 \rightarrow \nu_3+\gamma)&>&4\times
 10^{45}(\frac{2\sqrt{2}}{U_{\mu1}}\frac{\sqrt{2}}{U_{\mu3}})^2 
(\frac{3.2\times 10^{-2}}{m_1(\eV)})^5\quad{\rm years} \nonumber 
\eea 
where we have used the present limits obtained in Ref.\cite{Smirnov} 
\be 
\frac{f_{e\mu}^2}{\bar{M^2}}<7 \times 10^{-4}G_F \;. 
\label{femu} 
\ee 
\par
The amplitude for radiative decay of charged leptons also induced 
by Zee boson exchange at one loop level is 
\be
A(\mu \ra e+\gamma)=\frac{e}{768\pi^2{\bar M^2}}f_{\mu\tau}f_{e\tau}
{\bar u_e(p')}\sigma_{\mu\nu}q^\mu\e^\nu(1+\gamma_5)u_\mu(p) \;.
\ee
For $\tau \ra e \gamma$ and $\tau \ra \mu \gamma$ the parameters
$f_{\mu\tau}f_{e\tau}$ should be replaced with $-f_{\mu\tau}f_{e\mu}$ and $
f_{e\mu}f_{e\tau}$, respectively. 
The branching ratio is 
\be 
Br(\mu \ra e+\gamma)=\frac{\a}{3072\pi} 
(\frac{f_{\mu\tau}f_{e\tau}}{{\bar M^2}})^2\frac{1}{G_F^2}<4 \times
 10^{-27} 
\ee 
by using the upper limit (\ref{femu}) and Eqs.(\ref{f1},\ref{f2}). 
Thus the radiative decays of neutrinos and charged leptons, 
induced by the Zee boson, are negligible. 
\vskip 0.5cm 
\par 
In conclusion we have searched for solutions within the framework of 
the Zee mass matrix, that has vanishing or in general very small 
diagonal elements, by taking maximal mixing angle $\th_2\simeq \pi/4$ 
between $\nu_\mu$ and $\nu_\tau$. 
The solution we have found is given by Eq.(\ref{M3}). It 
corresponds to the bi-maximal solution, which requires large 
mixing angles for both solar- and atmospheric neutrinos. 
The two heaviest neutrinos $\nu_1$ and $\nu_2$, which are approximately 
degenerate, and the lightest neutrino $\nu_3$ are given by 
\bea 
\nu_1&\simeq&
 \frac{1}{\sqrt{2}}\nu_e-\frac{1}{2}\nu_\mu+\frac{1}{2}\nu_\tau 
\;, \nonumber\\ 
\nu_2&\simeq&
 \frac{1}{\sqrt{2}}\nu_e+\frac{1}{2}\nu_\mu-\frac{1}{2}\nu_\tau 
\;,\\ 
\nu_3&\simeq& \frac{1}{\sqrt{2}}\nu_\mu+\frac{1}{\sqrt{2}}\nu_\tau 
\nonumber \ . 
\eea 
The solution thus found requires a large hierarchy for the couplings of 
the Zee boson to leptons, in the form 
$f_{e\mu}\gg f_{e\tau} \gg f_{\mu\tau}$ 
in contrast to "natural" expectations. 
It is therefore desirable to impose an approximate $L_e-L_{\mu}-L_{\tau}$ 
symmetry in the Zee model in order to explain this hierarchy. 
The model will undergo severe tests in the future neutrino experiments. 
\vskip 0.5cm 
\noindent 
{\bf Acknowledgment}\\ 
One of the authors (M.M) expresses his thanks to 
the Royal Swedish Academy of Sciences (RSAS) 
for the support by a grant. M.T is thankful to the High Energy Group in 
CFIF/IST (Portugal) for their hospitality. His work is supported by the 
Grant-in-Aid for Science Research, Ministry 
of Education, Science and Culture, Japan (No.1014028, No.10640274).
\vskip 0.5cm     

\end{document}